\documentclass[runningheads]{llncs}

\usepackage{amssymb}
\usepackage{xspace}
\usepackage{amsmath}
\usepackage{xcolor}
\usepackage{graphicx}
\usepackage{multicol}
\usepackage{paralist}
\usepackage{comment}
\usepackage{enumitem}
\usepackage{wrapfig}
\usepackage{ifthen}
\newboolean{long}
\setboolean{long}{false}

\usepackage{bbm}
\usepackage{pgf,pgfarrows,pgfnodes}
\usepackage{tikz}
\usetikzlibrary{arrows,automata}
\usepackage{url}
\usepackage{algorithm,algorithmicx,algpseudocode}

\newcommand{\ecl}[1]{\mathit{ecl}(#1)}

\newcommand{\its}{\ensuremath{\mathcal{M}}}
\newcommand{\itss}{\ensuremath{\mathcal{M}^*}}
\newcommand{\itsp}{\ensuremath{\mathcal{M}^+}}
\newcommand{\defstyle}[1]{\textbf{#1}}
\newcommand{\cls}{\ensuremath{\mathcal{C}}}
\newcommand{\lang}{\ensuremath{\mathcal{L}}}
\newcommand{\fo}{\varphi}
\newcommand{\infor}{\Phi}
\newcommand{\sat}{\textsf{SAT}\xspace}

\newcommand{\val}{\textsf{VAL}\xspace}
\newcommand{\mc}{\textsf{MC}\xspace}
\newcommand{\mb}{\textsf{MB}\xspace}
\newcommand{\mcpm}{\textsf{MCPM}\xspace}
\newcommand{\mspm}{\textsf{MSPM}\xspace}
\newcommand{\epm}{\textsf{EPM}\xspace}
\newcommand{\sts}{\ensuremath{\mathsf{s}\xspace}}
\newcommand{\tableau}[1]{\ensuremath{\mathcal{#1}}}

\begin{document}
\sloppypar
\title{Model checking and model synthesis \\ from partial models: \\ a logic-based perspective}

\titlerunning{Partial model checking and synthesis}

\author{Valentin Goranko}

\authorrunning{Goranko}

\institute{Stockholm University, Sweden\\
\email{valentin.goranko@philosophy.su.se}\\
}

\maketitle

\begin{abstract}
I consider the following generic scenario: an abstract model M of some `real' system is only partially presented, or partially known to us, and we have to ensure that the actual system satisfies a given specification, formalised in some logical language. This scenario has at least two essentially different interpretations, leading to two, essentially different, formal logical and algorithmic problems: ``Model Synthesis from Partial Models'', where \emph{some} `admissible' extension of M to a full model must satisfy the specification, and ``Model Checking of Partial Models", where \emph{all} `admissible' extensions of M to a full model must satisfy the specification. These problems naturally extend the classical logical decision problems of Satisfiability, Validity, and Model Checking. 

Here I briefly discuss both problems in the contexts of classical, modal and temporal logics. I make some observations, state some open questions, and outline a general tableaux-style procedure that solves the problem of unconstrained model synthesis from finite partial models for several well-known modal and temporal logics, incl. K, LTL, CTL, ATL.
\end{abstract}

\section{Introduction}
\label{sec:intro} 

Consider the following generic scenario: an abstract model $\its$ of some `real' system is only partially presented, or only partially known  to us. Besides, it is known that the full model $\itss$ must belong to a given class of models $\cls$. We have to ensure that the actual system, represented by its full model $\itss$, satisfies a given specification $\Phi$, formalised in a suitable logical language. 

This scenario has at least two essentially different interpretations, leading to two, essentially different yet related, logical and algorithmic problems:

 \begin{enumerate}
\item 
\emph{The design/construction of the full system is yet to be completed}, and the given partial model $\its$ represents the current stage of its construction, presumably consisting of some necessary components, possibly related to its desired behaviour.  
 The problem now is a \emph{designer's task}, viz. to \emph{complete} the construction of $\its$ to a full model $\itss$ which belongs to the class $\cls$ and satisfies the prescribed specification $\Phi$.  I will call that task the problem of \defstyle{Model Synthesis from Partial Models} (\mspm). A weaker, non-constructive version of it, is the problem of \defstyle{Extendability of Partial Models} (\epm), which can be formally stated as a pure decision problem, only asking whether  \emph{some admissible extension/completion of the partial model $\its$ to a model in $\cls$ satisfying $\Phi$ exists}.

\item  \emph{The full system already exists out there, but is only partially known}, and the given partial model $\its$ represents what is known about that system. In addition, the full model $\itss$ is known to belong to the class $\cls$, possibly described using $\its$. In particular, the class $\cls$ describes how $\its$ may be related to $\itss$.  
The problem now is a \emph{verifier's task}: to check, only based on the partial model $\its$, that the full model $\itss$ satisfies the prescribed specification $\Phi$. Practically, that amounts to checking that \emph{every} possible (`admissible') completion of the partial model $\its$ to a full model $\itsp$ in the class $\cls$ satisfies the specification $\Phi$.  I will call that task the problem of  \defstyle{Model Checking of Partial Models} (\mcpm). It can be formally stated as a logical decision problem, asking \emph{whether all admissible completions of the partial model $\its$ to a model in $\cls$ satisfy $\Phi$}. 

A natural extension of this problem assumes that the agent can set up suitable experiments and make observations to \emph{learn} more and more from the model, until sufficient information is obtained to solve the \mcpm problem. 
This links that problem with  \emph{formal learning theory} \cite{Kelly1996}, \cite{sep-learning-formal}.   

\end{enumerate}

As shown further, the decision problems \mcpm and \epm together subsume the classical logical decision problems of satisfiability, validity, and model checking. Moreover, under common natural semantic assumptions,  \mcpm and \epm are dual to each other and inter-reducible, just like satisfiability and validity are. 

\smallskip
\textit{Related work.} 
Many instances of both problems have already been implicitly studied in various contexts in the literature on model checking and model synthesis. However, according to my knowledge, these problems have not been explicitly stated in such generic way so far, nor studied for their own sake, in the sense discussed here. Here I only provide a partial, and \emph{inevitably incomplete} list of references to some related works.  

\begin{itemize}
\item Conceptually closely related earlier works are:  
\cite{DBLP:conf/msras/Staruch04}, proposing the use of partial structures to model multi-agent systems where the agents possess only partial information, gradually completed as the knowledge grows, and 
\cite{DBLP:journals/sLogica/StaruchS05}, taking an algebraic approach to the completion of partial first-order models, again representing partial information about the actual world.

\item A technique called `partial model checking' was introduced in \cite{DBLP:conf/lics/Andersen95} (see also, \cite{DBLP:journals/sttt/AndersenL99}, 
\cite{DBLP:conf/tacas/CostaBBDG18}) for verifying concurrent systems by gradually removing concurrent components and then reducing the specification, until completely checked, 
in order to avoid the state explosion problem. This idea is only implicitly and technically related to the \mcpm problem discussed here.

\item A topic more closely related to the present work 
is \emph{synthesis of reactive programs and systems under incomplete information} \cite{Kupferman2000}, where an open system must be guaranteed to satisfy a given safety specification but can only partly read the input signals generated by its environment. 
An important practical case of synthesis from partial models is the problem of \emph{controller synthesis}. 
For lack of space, these links are only mentioned here, but they will be explored more properly in a follow-up full paper.

\item 
Another, more explicitly related to the present work, is the line of research in modal logic, involving modal operators that change the models dynamically, in the course of formula evaluation.  
It goes back, inter alia, to Renardel de Lavalette's \emph{dynamic modal logic DML}  
\cite{logcom/Lavalette04}, van Benthem's \emph{sabotage logics} \cite{DBLP:conf/birthday/Benthem05}, \cite{DBLP:journals/logcom/AucherBG18} and 
 Gabbay's \emph{reactive Kripke semantcs} 
 \cite{DBLP:series/cogtech/Gabbay13}. These were followed by various further proposals for \emph{model update logics}, including enrichments of modal logics with \emph{global and local graph modifiers}  
\cite{DBLP:journals/entcs/AucherBCH09}, and \emph{relation-changing modal operators and logics} \cite{DBLP:journals/igpl/ArecesFH15}. 
Typically, these approaches involve operations on models that not only extend, but also shrink or modify them in various ways and, moreover, add syntactic instructions for such operations in the language. These features can often lead to undecidability of both model checking and satisfiability of such logics, see \cite{DBLP:conf/tableaux/ArecesFHM17}, \cite{DBLP:journals/logcom/ArecesFHM18}.

\item Other relevant references to mention here include \cite{ictac/BrenasES16} and  \cite{icdt/ItzhakyKRSTVZ17}. 
\end{itemize}

\section{Model checking and extendability from partial models as logical decision problems}
\label{sec:logicaldecision} 

Here I briefly discuss EPM and MCPM in the context of the three classical logical decision problems, all generically defined for a given logical language $\lang$ with specified semantics in terms of truth (satisfaction)  and validity in a model (at least over a given class of models $\cls$):  

\begin{itemize}
\item \textbf{Satisfiability} (\sat): given a formula $\fo$ of $\lang$, to determine whether $\fo$ is satisfied in some model (in $\cls$). 

\item \textbf{Validity} (\val): given a formula $\fo$ of $\lang$, to determine whether $\fo$ is valid in every model (in $\cls$).  

\item \textbf{Model checking} (\mc): given a formula $\fo$ of $\lang$, 
a model $\its \in \cls$, and (in the case of locally defined notion of truth) a state $\sts \in \its$, to determine whether $\fo$ is true/satisfied (at $\sts$) in the model $\its$.  
\end{itemize}

Some important remarks are in order: 

\smallskip
1. In the special case when the given partial model $\its$ is empty and $\cls$ is the class of all models for $\lang$, the problem \mcpm reduces to \val. More generally, \mcpm subsumes  checking validity in the given class $\cls$, hereafter denoted $\val_\cls$.   

\smallskip
2. The 
problem \mcpm also subsumes \mc as a special case when the given partial model $\its$ is complete (i.e., no extension is needed or allowed).   

\smallskip
3. In the special case when the given partial model $\its$ is empty and $\cls$ is the class of all models for $\lang$, the problem \epm reduces to \sat. 
 More generally, it subsumes checking satisfiability in the class $\cls$, hereafter denoted $\sat_\cls$.  

Respectively, the problem \mspm reduces to the problem of \emph{Constructive Satisfiability}, aka \emph{Model Building} (in the given class $\cls$), hereafter denoted \mb, when the given partial model $\its$ is empty. (Unlike the usual decision problem \sat, where only existence of a model is to be determined, the Constructive Satisfiability asks for constructing a satisfying model whenever one exists.)  

\smallskip
4. The problem \mspm subsumes \mc, too, when the given partial model $\its$ is complete (i.e., no extension is needed or allowed).   

\smallskip
5. Just like \val and \sat, the problems \mcpm and \epm  are usually dual to each other:  
the output of \mcpm with input formula $\fo$, model $\its$ and (if applicable) state $\sts \in \its$  is the reverse output (swapping `yes' and 'no') of \mcpm with input formula $\lnot \fo$, model $\its$ and state $\sts$. It is therefore sufficient to develop decision methods for either of these. Moreover, all usual satisfiability decision procedures are constructive, so any natural decision method for  \epm should be constructive and provide a sufficient information for building a satisfying full model extending the initial partial model $\its$, whenever one exists.  

\medskip
Thus, each of \mcpm and \epm naturally extends the classical logical decision problems of \sat, \val and \mc. The scheme on the left below puts all these together, while the one on the right relates the respective constructive versions \mb and \mspm to \sat and \epm, where the arrows indicate subsumption. 

 \begin{center} 
 \begin{tikzpicture}[->,>=stealth',shorten >=1pt,
node distance=2.2cm,
                   thick]

  \tikzstyle{every state}=[fill=white,draw=black,text=red]

  \node[state]         (1)                    {\val};
   \node[state]        (2) [above of=1] {\mcpm};    
  \node[state]         (3) [above right of=2] {\mc}; 
  \node[state]         (4) [below right of=3] {\ \ \epm \ ~};   
  \node[state]         (5) [below of=4] {\sat};

  \path (3) edge             node [left] {} (2)     
           (3)  edge              node {} (4)                            
           (1)    edge              node {} (2)
           (5)    edge              node {} (4);                   

\draw [-, dashed] (2) to  node [auto, swap] {duality} (4);     
\draw [-, dashed] (1) to  node [auto, swap] {duality} (5);                            
\end{tikzpicture}
\ \ \ \ \ \ \ \ 
 \begin{tikzpicture}[->,>=stealth',shorten >=1pt,
node distance=2.2cm,
                   thick]

  \tikzstyle{every state}=[fill=white,draw=black,text=red]

  \node[state]         (1)                    {\val};
   \node[state]        (2) [above of=1] {\mcpm};    
  \node[state]         (3) [above right of=2] {\mc}; 
  \node[state]         (4) [below right of=3] {\mspm};   
  \node[state]         (5) [below of=4] {\mb};

  \path (3) edge             node [left] {} (2)     
           (3)  edge              node {} (4)                            
           (1)    edge              node {} (2)
           (5)    edge              node {} (4);

\draw [-, dashed] (4) to  node [auto, swap] {\small{constructive}} (2);     
\draw [-, dashed] (2) to  node [auto, swap] {\small{duality}} (4);     
\draw [-, dashed] (5) to  node [auto, swap] {\small{constructive}} (1);     
\draw [-, dashed] (1) to  node [auto, swap] {duality} (5);                            
\end{tikzpicture}
\end{center}

\section{Some special cases} 
\label{sec:cases}

Hereafter, I will focus mainly on the \epm and \mspm problems, which I will discuss briefly for some of the most commonly used logical languages, for purposes where these problems make particularly good sense.

\subsection{First-order logic} 
\label{sec:FOcase}

The problems of model checking and model synthesis of partial models naturally apply in first-order logic (FOL). As mentioned earlier, a relevant study here is \cite{DBLP:journals/sLogica/StaruchS05}. Here I will only add some immediate observations on this case. 
For technical ease, I will consider first-order languages with relational signature only, including equality, and possibly constant symbols. 

\begin{itemize}
\smallskip
\item When the given partial model $\its$ is finite, it can be described by a single FO sentence, viz the existential closure of its atomic diagram $ExDiag_{\its}$. 
Since every model of this sentence contains an isomorphic copy of $\its$, solving the \epm problem for $\its$ and a specification $\Phi$, over the class of all FO structures, amounts to solving the \sat problem for $\Phi \land ExDiag_{\its}$, thus making \epm and \sat mutually reducible here. Likewise for \mb and \mspm. Thus, all these problems are, generally, only semi-decidable. 
This observation can be modified for various special types of extensions, e.g. when the given partial model must be an induced substructure of the desired extension. 

\smallskip
\item The case of an infinite given partial model is more complex. Mathematically, the \epm problem can still be reduced to \sat, but for infinite theories rather than single sentences. To tackle that problem algorithmically, the infinite partial model must be \emph{finitely presentable}. That opens a rich line of research, which will not be discussed here but followed up in a further work. 
\end{itemize}

\subsection{Modal and temporal logics} 
\label{sec:MLcase}

Typically, modal and temporal languages are used for describing the behaviour of various state transition systems. A variety of natural extensions of the given partial model can be defined in this context, including: 
\begin{itemize}
\smallskip
\item extensions where the state space remains fixed, but only transitions may be added, or only more atomic propositions (presumably, occurring in the specification $\Phi$) can be additionally interpreted in the extension, or both. If the given partial model is finite, the problems \epm and \mcpm are clearly decidable, but their complexity is  generally a subject of investigation. The case of infinite initial partial models seems mostly unexplored yet.   

\smallskip
\item extensions where, along with the transition relation and the valuations of the atomic propositions, the state space can expand, too. A special case is where the initial partial model $\its$ must remain an induced subgraph of the completed model, i.e. no new transitions are allowed between states of $\its$. 
 All these cases generate \epm and \mcpm problems that may be undecidable, even when the given partial model is finite, but they can often be reduced to decidable \sat problems in suitably enriched hybrid modal languages. 
\end{itemize}

\subsection{Logics for multi-agent systems} 
\label{sec:MALcase}

Multi-modal languages explicitly involving agents in the language allow for an even richer variety of natural notions of model extensions, in terms of which the decision and construction problems discussed here can be considered. Besides all types of extensions of interpreted transition systems mentioned above, some 
agent-related extensions naturally emerge, too, including:      

\begin{itemize}
\item extensions preserving the existing agents but enriching their action sets, 
thus enhancing their functionalities and enabling new transitions in the system. 

\item truly \emph{dynamic extensions}, also involving addition of new agents with their capacities to affect the behaviour of the system. 
An important particular case of practical application of the \mspm problem, is the \emph{controller synthesis} problem, where the desired controller can be regarded as a special new agent whose task is to guarantee the satisfaction of a given safety specification. 
\end{itemize}

\section{A generic tableaux-based method for model synthesis from partial models in modal and temporal logics}
\label{sec:tableauMC}

The traditional tableaux-based procedures for testing satisfiability attempt to build a satisfying model for the input formula from scratch. Many of these procedures can be smoothly modified to start from a given partial model and attempt to extend it in a prescribed way, to a full model from the target class $\cls$, satisfying the input formula. Here I will outline such a generic tableaux-based procedure, applicable to a wide variety of modal, temporal, and multi-agent logics, including K, LTL, CTL, ATL (cf e.g.  \cite{TLCSbook} on these), that solves the essentially unconstrained \mspm problem, for the class $\cls$ of all models of the respective logic that extend $\its$ in an explicitly specified way (see below).

I will follow the incremental tableau-building methodology presented and applied in \cite[Ch.13]{TLCSbook} to LTL and CTL (see also \cite{GorankoShkatov09ToCL},  \cite{igpl/AjspurGS13}, \cite{tocl/Cerrito0G15},  
for tableaux systems for some multi-agent logics), to outline a generic procedure for model synthesis from partial models, applicable, \emph{mutatis mutandis}, to a variety of logics, including all those mentioned above. 
Here I will assume that the admissible extensions may be obtained by adding more states and transitions, but no atomic propositions to labels of the existing states, nor new transitions between existing states. The method is amendable to various modifications, reflecting other types of admissible extensions, too. 
Details will be provided in a follow-up full paper. 

The tableau procedure presented below takes as input a triple $(\infor, \its, \sts)$ consisting of a 
given formula $\infor$, a finite partial model $\its$, and a designated root state $\sts$. It attempts to construct  a non-empty graph $\tableau{T}^{\infor}$, called a \defstyle{tableau}, representing (in a way) sufficiently many possible `quasi-models' (Hintikka structures) for $\infor$, $\its$ and $\sts$ from which a desired model can be extracted, if there is any.  
The procedure consists of three major phases (cf.  \cite[Ch.13]{TLCSbook} for details):
\begin{enumerate}
\item 
In the \defstyle{construction  phase}, a finite directed graph $\tableau{P}^{\infor}$ with labeled nodes, called the \defstyle{pretableau} for $\infor$, is produced, following prescribed \defstyle{construction rules}. 
The pretableau has two types of nodes: \defstyle{states} and \defstyle{prestates}. The states are labeled with so called `fully expanded' subsets of the extended closure $\ecl{\infor}$ and represent states of a Hintikka structure from which the model  will be eventually built, while the prestates can be labeled with any subsets of $\ecl{\infor}$ and they only play an auxiliary and temporary role.
The special aspect of the construction phase now is that it starts not with a single initial prestate, but with a copy of the partial model $\its$, each state of which represents a prestate in the tableau. These prestates are expanded further, step-by-step, by fully expanding their labels, adding respective offspring states and necessary successor states until saturation, as in the standard tableau construction.

\smallskip
\item  
In the \defstyle{prestate elimination phase}, all prestates are removed just like in the standard construction, thus producing the \defstyle{initial tableau} for $(\infor, \its, \sts)$. 

\smallskip
\item 
The \defstyle{state elimination phase} removes, using \defstyle{state elimination rules}, all (if any) `bad' states from the initial tableau that cannot be satisfied in any full  model of the desired type. 
Removal of a state can be done for two reasons, viz.: (i) some of the successor states that the state needs have already been removed in the elimination process so far, and (ii) explicit eventualities in the label of the state are not satisfied in the current tableau. 
    The state elimination phase eventually results in a (possibly empty) subgraph
$\tableau{T}^{\infor}$ of the initial tableau, called the \defstyle{final tableau} for $\infor$.
The special aspect of the elimination phase now is that every state $t$ in the input partial model $\its$ must retain at least one representing state in the tableau at every stage of the elimination. Thus, if all representing states of $t$ are to be removed, the elimination phase ends and the resulting final tableau is declared closed.  
\end{enumerate}

\smallskip
After the end of the elimination phase, if the final tableau $\tableau{T}^{\infor}$ is not declared closed yet, 
and if there is a state in $\tableau{T}^{\infor}$ representing the state $\sts$ in $\its$ and containing $\infor$ in its label, the tableau is declared \defstyle{open} and the input formula $\infor$ is pronounced satisfiable at $\sts$ in some admissible extension of $\its$; otherwise, the tableau is declared \defstyle{closed} and $\infor$ is claimed not satisfiable in any such extension. 

\smallskip
The tableau-like model building procedure described above is generically applicable to a wide range of modal and temporal logics, and solves the \mspm problem for them, as stated in the next general result. 

\begin{theorem}[Soundness and completeness] 
\label{thm:compl}
For each of the logics \textbf{K}, \textbf{LTL}, \textbf{CTL}, \textbf{ATL},  
for any input $(\infor, \its, \sts)$, the final tableau $\tableau{T}^{\infor}$ for that logic is open iff 
$\infor$ is satisfiable at $\sts$ in some admissible extension of $\its$. 
\end{theorem}

\end{document}